# Hall Effect anomalies near Quantum Critical Point in $CeCu_{6-x}Au_x$.

N. E. Sluchanko[1*], V. V. Glushkov[1,2], S. V. Demishev[1,2], N. A. Samarin[1],

G. S. Burhanov[3], O. D. Chistiakov[3], D. N. Sluchanko[1].

[1] *A. M. Prokhorov General Physics Institute of RAS, 38, Vavilov str., Moscow, 119991, Russia*

[2] *Moscow Institute of Physics and Technology, 9, Institutskii per., Dolgoprudny,*

*Moscow region, 141700, Russia*

[3] *A. A. Baikov Institute of Metallurgy and Materials Technology of RAS,*

*49, Leninskii pr., Moscow, 119991, Russia*

*-E-mail: nes@lt.gpi.ru

Precision measurements of the Hall effect have been carried out for both archetypal heavy fermion compound - $CeCu_6$ and exemplary solid solutions $CeCu_{6-x}Au_x$ ($x$= 0.1 and 0.2) with quantum critical behavior. The experimental results have been obtained by technique with a sample rotation in magnetic field in the temperature range 1.8-300K. The experiment revealed a complex activation type dependence of the Hall coefficient $R_H(T)$ in $CeCu_6$ with activation energies $E_{a1}/k_B \approx 110K$ and $E_{a2}/k_B \approx 1.5K$ in temperature ranges 50÷300K and 3÷10K, respectively. Microscopic parameters of charge carriers transport (effective masses, relaxation time) and localization radii $a_{p1,2}^*$ of heavy fermions ($a_{p1}^*$(T>50K)~ 1.7 Å and $a_{p2}^*$(T<20K)~ 14 Å) were estimated for $CeCu_6$. The second angular harmonic contribution has been established in the Hall voltage of $CeCu_{5.9}Au_{0.1}$ and $CeCu_6$ at temperatures below $T^*$~24K. A hyperbolic type divergence of the second harmonic term in Hall effect $R_{H2}(T) \approx C(1/T-1/T^*)$ at low temperatures is found to be accompanied with a power-law behavior $R_H(T)$~ $T^{-0.4}$ of the main contribution in the Hall coefficient for $CeCu_{5.9}Au_{0.1}$ compound with quantum critical behavior.

**1.** During last decade it was demonstrated perfectly [1-7] that $CeCu_{6-x}Au_x$ solid solutions are in the number of the most convenient and interesting examples of the systems with non-Fermi-liquid behavior in the vicinity of antiferromagnetic quantum critical point (QCP). Among the $CeCu_{6-x}Au_x$ family the $CeCu_6$ is the canonical heavy fermion (HF) intermetallic compound with an enhanced Pauli magnetic susceptibility and a huge value of the low temperature electronic specific heat coefficient $\gamma$ = 1.53 J/(mole K$^2$) [8]. These features place this compound together with $CeAl_3$ [9] as the heaviest of all known HF systems. Alloying with Au induces the long-range antiferromagnetic order above QCP at $x_c \approx 0.1$ with the Neel temperature that increases linearly in the $0.1 < x \leq 1$ range and reaches $T_N \approx 2.3K$ for $CeCu_5Au$ [10]. Understanding the behavior of this kind of systems close to antiferromagnetic QCP is a current area of active research. In this field, the main question is the mechanism by which the mass of heavy electrons diverges in the approach to the antiferromagnetic instability [11].

According to the comments proposed in [11], Hall coefficient is the one of the most promising but, surprisingly, untested probes of the quasiparticles involved in the conduction process on both the paramagnetic and antiferromagnetic sides of the QCP. From the point of view of [11], the Hall coefficient variation through the quantum phase transition (QPT) can be considered as a key experiment to choose between different scenario of QPT and to test the mechanism by which the Fermi surface transforms between the paramagnet and antiferromagnet.

Up to now the Hall effect measurements of $CeCu_6$ [12-15] and the exemplary solid solutions $CeCu_{6-x}Au_x$ [16-17] have been produced by ordinary ac- (or dc-) four probe method where the Hall resistivity $\rho_H(H,T)$ was deduced as a difference between two signals obtained for two opposite orientations of external magnetic field ***H*** which was perpendicular both to the sample surface and to the current ***I*** direction. However, the low dimensional character of the spin fluctuation spectra near the QCP $x_c \approx 0.1$ is well established for $CeCu_{6-x}Au_x$ at present [3-5]. Consequently, an appearance of an "induced anisotropy effects" can be also expected for the charge transport characteristics and, particularly, for the Hall coefficient behavior of $CeCu_{6-x}Au_x$.

2. The aim of this study was to carry out for the first time the comprehensive measurements of the Hall resistivity angular dependencies of $CeCu_{6-x}Au_x$ compounds in the nearest vicinity of the quantum critical point (*x=0, 0.1, 0.2*). The investigations have been carried out for a wide range of temperatures 1.8-300K in magnetic field of permanent magnets ($H_0 \sim 1-4\ kOe$). The experiments were performed with the help of the original installation using the step-by-step fixation of the sample rotated in steady magnetic field that was applied perpendicular to the rotation axis [18]. Special attention was paid to the symmetrical arrangement of the Hall voltage probes in the Hall effect measurements, because of the strong effect of negative magnetoresistance in $CeCu_{6-x}Au_x$ at temperatures T < 30K [19] (see also $\rho(T, H=70\ kOe)$ curves in Fig. 1).

The $CeCu_{6-x}Au_x$ high quality polycrystalline samples were synthesized from stoichiometric amounts of high purity components in an electric arc furnace with a non-consumable tungsten crucible on a water-cooled cold copper finger under a high-purity helium atmosphere. The composition homogeneity was attained in the bulk materials by a repeated arc melting of the starting components in the stoichiometric ratio with subsequent annealing in evacuated quartz tubes. The samples' characterization methods (room temperature X-ray powder diffraction and microprobe analysis experiments) have been applied to identify the $CeCu_{6-x}Au_x$ solid solutions under investigation as essentially single-phase compounds. The orthorhombic structure parameters of $CeCu_{6-x}Au_x$ alloys are performed in the Table 1. Note that $CeCu_6$ undergoes a phase transition from the orthorhombic structure to the monoclinic one below $T_S \approx 200K$ with a small distortion of about 1.5° [20]. For comparison, the structural data from Ref. 20 are also included in the Table 1.

3. The results of resistivity measurements for the studied samples are shown in Fig. 1. From the data of Fig.1 the low temperature resistivity $\rho(T)$ maximum is clearly detected at $T_{max}^\rho \approx 12.5K$, 11K and 5.5K for x=0, 0.1 and 0.2 respectively. It is clearly identified that the $T_{max}^\rho$ value exceeds noticeably the Kondo temperature $T_K \approx 5\text{-}6K$ estimated for $CeCu_6$ and $CeCu_{5.9}Au_{0.1}$ in [21-22]. The issue of $T_{max}^\rho >> T_K$ is traditionally associated with an essential contribution from inelastic Kondo-scattering processes in charge transport of $CeCu_{6-x}Au_x$. In magnetic field $H \sim H_K = k_B T_K/\mu_B \approx 70$ kOe the low temperature resistivity maximum at $T_{max}^\rho$ can be evidently detected at $T_{max}^\rho \approx 15 \div 20K >> T_K$ on the $\rho(T)$ curves for all studied $CeCu_{6-x}Au_x$ compounds (Fig. 1).

Then, considering Hall effect measurements in $CeCu_{6-x}Au_x$ it is worth to note that only small enough external magnetic field intensities H<5 kOe have been used in this study to avoid any influence of transverse magnetoresistance component in the Hall signal. As it was shown previously in [19], at low temperatures $T \leq 30K$ the negative magnetoresistance (MR) behavior in magnetic field is quadratic $\rho(H) \sim H^2$ in $H \leq 30$ kOe range for $CeCu_{6-x}Au_x$ ($x \leq 0.2$), so, the MR component measured from the Hall probes is negligible below H~5 kOe.

The families of angular dependencies of the Hall resistance for $CeCu_6$, $CeCu_{5.8}Au_{0.2}$ and $CeCu_{5.9}Au_{0.1}$ are shown in Figs. 2a, 2b and 3 respectively. The data were obtained for temperatures between 1.8K and 300K in magnetic field $H_0 \approx$ 3.7 kOe. It is natural to expect for $CeCu_{6-x}Au_x$ solid solutions a simple dependence $\rho_1 \sim \rho_{H1}\cos\varphi$ of the Hall signal which is caused by a variation of the amplitude of a scalar product (***n H***) (***n*** – normal to the sample surface) when rotating the sample in steady magnetic field [18]. In the temperature interval 1.8-300K applied in this research, all studied $CeCu_{6-x}Au_x$ compounds are paramagnets. So, it is difficult to expect an appearance of induced magnetic anisotropy and, as a sequence, of a complicated and unconventional angular dependencies of charge transport parameters (see, e.g., [18]) in these polycrystalline samples. Indeed, the cosinusoidal dependencies $\rho_1 \sim \rho_{H1}\cos\varphi$ have been recorded for $CeCu_{5.8}Au_{0.2}$ in the whole temperature range 1.8-300K (Fig. 2b), whereas a second harmonic component $\rho_2 \sim \rho_{H2}\cos(2\varphi)$ has been detected additionally to the main contribution both for $CeCu_6$ and $CeCu_{5.9}Au_{0.1}$ at low temperatures (Fig. 2a and Fig. 3). It should be emphasized that the above mentioned $\rho_{H2}$ component in the Hall signal of $CeCu_{5.9}Au_{0.1}$ causes a noticeable changes of the extremuma positions on the experimental angular dependencies $\rho_H^{exp}(\varphi)$ at liquid helium temperatures (see, e.g., curve $T$=2K in Fig. 3). In this situation the traditional technique applied to the low temperature Hall effect measurements of $CeCu_{6-x}Au_x$ [12-17] leads evidently to incorrect values of the Hall coefficient for these systems with quantum critical behavior.

4. The Hall resistivity angular dependencies (Fig. 2-3) were analyzed within relation

$$\rho_H(\varphi, T_0, H_0) = \rho_{H0} + \rho_{H1}\cos\varphi + \rho_{H2}\cos(2(\varphi-\Delta\varphi)) \qquad (1),$$

which includes the second harmonic contribution $\rho_{H2}$ with the phase shift $\Delta\varphi$ in addition to the main component $\rho_{H1}$ (odd with respect to the magnetic field) and the angle independent shift $\rho_{H0}$. The procedure for separating the contributions in (1) is most clearly shown in Fig. 3, where the components with the amplitudes $\rho_{H1}$ and $\rho_{H2}$ are plotted along with the experimental curve $\rho_H^{exp}(\varphi, H_0 \approx$ 3.7 kOe) for various temperatures in 1.8-20K range. The approximation of the experimental data by (1) (solid lines in Fig.3) can be considered as a good fit for the whole temperature range 1.8-300K used in this study. Then, the amplitudes of the Hall resistivity components $\rho_{H1}(T,H_0)$ and $\rho_{H2}(T,H_0)$ were used to plot the temperature dependencies of the Hall coefficient $R_H(T,H_0 \approx$ 3.7 kOe) (Fig. 4) and the second harmonic term $R_{H2}(T, H_0 \approx$ 3.7 kOe) (Fig. 5) in $CeCu_{6-x}Au_x$ solid solutions. It is important to note that according to the results obtained in [12-14] for polycrystalline samples of $CeCu_6$ the $R_H(T)$ stays positive above liquid helium temperature and is characterized by a wide maximum of prominent amplitude at $T_{max}^{R_H} \approx T_{max}^{\rho} \sim 10K$. Moreover, the transition region (interval $T \leq T_{max}^{\rho} \sim 10K$) from incoherent to coherent scattering is complicated for $CeCu_{6-x}Au_x$ and happens to be very sensitive to the impurities and partial substitution for Ce and Cu [12,13]. Therefore, the Hall effect analysis developed in this study is concentrated only on *(i)* the intermediate temperature range ($T \geq 10K$) dependencies of $R_{Hi}(T)$ and *(ii)* the comparison of the behavior of the $R_{Hi}(T)$ components for these compounds near QCP with various substitution of Au and for canonical heavy fermion system $CeAl_3$ at temperatures 1.8-300K.

The temperature dependencies of $R_H(T)$ for various $x$ in $CeCu_{6-x}Au_x$ alloys are slightly different in magnitude of the main component in the Hall effect, but they are characterized by an absolutely different analytical behavior (Fig. 4). Indeed, in the temperature intervals 50-300K (I) and 3-10K (II) the $R_H(T)$ curve of $CeCu_6$ can be well approximated by a complicated activation type behavior

$$R_H(T) \sim \exp(E_{a1,2}/k_B T) \qquad (2)$$

which is similar to that found very recently for $CeAl_3$ [23]. The activation energy values $E_{a1}/k_B \approx 110 \pm 8K$ and $E_{a2}/k_B \approx 1.5 \pm 0.1K$ are deduced from the reciprocal logarithmic plot of the data for $CeCu_6$ Fig. 6 in I and II intervals respectively.

As it was mentioned in [18, 23], the activation type behavior of $R_H(T)$ in I and II ranges is very unusual for a metallic system and does not have an explanation within Kondo lattice model and the skew-scattering mechanism of charge transport [24-25]. Indeed, in the approach developed in [24-25] the spin-flip resonance scattering of conduction electrons on the localized magnetic moments of rare-earth ions is considered as the most important factor of charge transport. As a result, the anomalous positive Hall effect in compounds with heavy fermions as well as the low temperature resistivity anomalies (see Fig. 1) both seem to be exclusively due to the specific character of the scattering effects [24-25].

In our opinion, the $E_{a1,2}$ parameters defined for $CeCu_6$ from the data of Fig. 6, evidently like as for $CeAl_2$ [18] and $CeAl_3$ [23], can be regarded as a bound energies of two types' manybody states, which are formed in vicinity of *Ce 4f*-centers (i) when the excited levels of the Ce $^2F_{5/2}$-state are considerably populated (50÷300K, range (I) in Fig. 6), and (ii) when the electronic density fluctuations occur as fast transitions between the lowest doublet of $^2F_{5/2}$-state and conduction band (3÷10K, range (II) in Fig. 6). In this connection, the inset of Fig. 4 presents the scheme of crystalline electric field splitting of cerium $^2F_{5/2}$-state, as obtained from the results of inelastic neutron scattering experiments [26]. Thus, at intermediate temperatures 50÷300K the manybody states in $CeCu_6$ matrix are formed in collaboration with inelastic processes. In this temperature range the fast spin/charge fluctuations, which cause the charge carrier polarization, occur between the populated *4f*-band doublets and the conduction band states. According to [26], the excited doublets of cerium $^2F_{5/2}$-state prove to be considerably broadened and form above the main doublet a band of excited states of the width Γ, which is compared with the splitting value $Γ/2 ≈ E_{a1,2}/k_B ≈ 110K ~ Δ_1, Δ_2 ≈ 64-128K$ (see inset in Fig. 4).

The value of quasi-elastic peak width $Γ_0(T)/2≈6K$ obtained for $CeCu_6$ in the neutron scattering experiments [26] can be used within simple relation

$$Γ_0(T)/2 = \hbar / τ_{eff}(T) \quad (3)$$

to evaluate the relaxation time $τ_{eff}(T)$ and further on, taking into account the formula

$$m^*(T) = e\, τ_{eff}(T) / μ_H(T) \quad (4),$$

the data allow estimating the value and character of the temperature dependence of charge carrier effective mass. The *m\** values, estimated in the current approximation, are $m^*_1(T_0=77K) ≈ 137\, m_0$ and $m^*_2(T_0=8K) ≈ 145\, m_0$. These values are in a concordance with the early defined effective masses of spin-polaron and exciton-polaron states in strongly correlated electron systems - *FeSi* ($m^* ≈ 20÷90\, m_0$ [27]), $SmB_6$ ($m^* ≈ 20÷40\, m_0$ [28]), $CeAl_2$ ($m^*_{1,2} ≈ 55÷90\, m_0$ [28]) and $CeAl_3$ ($m^*_{1,2} ≈ 45÷90\, m_0$ [23]). The above deduced parameters $E_{a1,2}$ and $m^*_{1,2}$ can be applied also to estimate the manybody states' localization radii

$$a^*_{p1,2} = \hbar /(2E_{a1,2}m^*_{1,2})^{1/2} \quad (5).$$

For heavy fermions in $CeCu_6$ matrix this estimation leads to 1.7Å and 14Å values in the temperature ranges I and II, respectively. It should be pointed out that the small localization radius $a^*_{p1}$ (spatial size of nanoregions $a^*_{p1}≈1.7$Å $<a$, where *a* is the lattice constant, Table 1) corresponds to the "deep potential well" around the $Ce^{3+}$-sites ($E_{a1}/k_B ≈ 110±10K$, range I in Fig. 6), whereas at low temperatures the manybody states spread significantly ($a^*_{p2}≈14$Å) together with a dramatic decrease in their bound energy ($E_{a2}/k_B ≈ 1.5±0.1K$, range II in Fig. 6). It is obvious that such a small characteristic size of manybody states in $CeCu_6$ $a^*_{p1}≈1.7$Å cannot be explained within the Kondo lattice model. On the other hand, it is natural to assume that the dramatic increase of the manybody states' localization radius causes a further transition to a coherent state at low temperatures, which is accompanied by abrupt decrease of resistivity in rare-earth intermetallides with strong electron correlations.

It is necessary to mention also a significant difference in the Hall effect of the archetypal and very similar strongly correlated electron systems $CeCu_6$ and $CeAl_3$. Indeed, the appearance of the second harmonic contribution in the Hall signal was detected in this study for $CeCu_6$, contrary to the usual cosinusoidal angular dependencies that were recorded in [23] for $CeAl_3$. Moreover, the reduction of the low temperature bound energy $E_{a2}/k_B \approx 1.5K$ of many-body states in $CeCu_6$ in comparison with the value $E_{a2}/k_B \approx 3.3K$ detected for $CeAl_3$ can be considered, in our point of view, as a transition from activation to a wide range (1.8-30K) exponential dependence of the Hall coefficient

$$R_H(T) \sim T^{-\alpha}, \qquad \alpha = 0.41 \pm 0.02 \qquad (6),$$

that was found in present study for $CeCu_{5.9}Au_{0.1}$ compound with quantum critical behavior (Fig. 4). Then, moving to the antiferromagnetic side from QCP at $x \approx 0.1$, an essential lowering of the Hall coefficient values has been established for $CeCu_{5.8}Au_{0.2}$ (Fig. 4) in combination with disappearance of the second harmonic component in the Hall effect (Fig. 5). As a result, the QCP at $x \approx 0.1$ in $CeCu_{6-x}Au_x$ alloys proves to be distinguished by (i) the largest values of the Hall coefficient components $R_H(T)$ and $R_{H2}(T)$ (Figs. 4 and 5) and (ii) a power-law temperature dependence (6) at low temperatures 1.8-30K.

When analyzing a behavior of the second harmonic component $R_{H2}(T)$ in $CeCu_6$ and $CeCu_{5.9}Au_{0.1}$ the hyperbolic type divergence

$$R_{H2}(T) \approx C(1/T - 1/T^*) \qquad (7)$$

has been found in this study (Fig. 5). The value $T^* \approx 24K$ was deduced from the data of Fig. 5 both for $x=0$ and $x=0.1$ Au contents in $CeCu_{6-x}Au_x$ solid solutions. Additionally, a phase shift $\Delta\varphi$ between the main component and the second harmonic contribution has been detected from the analysis in terms of (1) from the data obtained for $CeCu_6$ and $CeCu_{5.9}Au_{0.1}$ (inset in Fig. 7).

Finally, to compare the results of present study with predictions of skew-scattering model [24-25], the charge carrier mobility $\mu_H(T) = R_H(T)/\rho(T)$ has been detected from the data of Fig.1 and Fig. 4. The family of the mobility $\mu_H(T)$ temperature dependencies obtained for $CeCu_{6-x}Au_x$ and $CeAl_3$ intermetallic compounds is shown in Fig. 7. The small enough values of $\mu_H(T) < 100$ cm$^2$/(V sec) are typical for these strongly correlated electron systems with fast spin/charge electron density fluctuations. From the double logarithmic $\mu_H(T)$ data presentation one may evidently conclude in favor of exponential type behavior of the charge carrier mobility

$$\mu_H(T) \sim T^{-\alpha} \qquad (8)$$

at temperatures below $T^* \sim 24K$ for all compounds placed on paramagnetic side of QCP. The exponent $\alpha$ values are in the interval 0.35-0.41 and increase slightly from $\alpha = 0.35$ in the canonical heavy fermion system $CeAl_3$ through $\alpha = 0.38$ in $CeCu_6$ to $\alpha = 0.41$ value detected for $CeCu_{5.9}Au_{0.1}$ compound with quantum critical behavior (Fig. 7). From the antiferromagnetic side of QCP a quasi-exponential $\mu_H(T)$ asymptotics is observed at intermediate temperatures 60-300K with $\alpha \approx 4/3$ and additionally this kind of behavior may be discussed in 1.8-10K range, where $\alpha \approx 0.16 \pm 0.01$ value has been deduced from the data of Fig. 7. At the same time, the Curie-Weiss type behavior of $\mu_H^{-1}(T) \sim (T - \Theta_p) \sim \chi^{-1}(T)$ predicted in [24, 25] for these systems within the skew-scattering model has not been detected in present study.

To summarize, the detailed measurements of the Hall effect in the exemplary $CeCu_{6-x}Au_x$ solid solutions performed in this study allowed us to separate and classify the contributions to the anomalous Hall effect in these compounds with heavy fermions and quantum critical behavior. For $CeCu_6$ and $CeCu_{5.9}Au_{0.1}$ the emergence of the second harmonic contribution was established in the angular dependencies of Hall signal. A hyperbolic type divergence of an amplitude of this even contribution in Hall effect at low temperatures T< T*~ 24K is found to be accompanied with a power-law behavior of Hall coefficient $R_H(T) \sim T^{-0.4}$ for $CeCu_{5.9}Au_{0.1}$ compounds with quantum critical behavior. A complex activated dependence of $R_H(T)$ has been observed in archetypal heavy fermion compound $CeCu_6$ in the temperature intervals 50-300K and 3-10K with activation energies $E_{a1}/k_B \approx 110K$ and $E_{a2}/k_B \approx 1.5K$ respectively, and microscopic parameters (effective masses and localization radii) of charge carriers have been deduced from the data obtained. The anomalies of transport characteristics observed in this study do not fit in with the prediction of the skew-scattering models, but can be interpreted in terms of spin-polaron manybody states in the matrix of $CeCu_{6-x}Au_x$ solid solutions.

This work was supported by the Russian Foundation for Basic Research (project no. 04-02-16721) and INTAS (project no. 03-51-3036) and by the Programs "Strongly Correlated Electrons in Semiconductors, Metals, Superconductors, and Magnetic Materials" of Russian Academy of Sciences and "Development of Scientific Potential" of Education and Science Ministry of RF. V.V.G. acknowledges support from Russian Science Support Foundation.

**References.**

1. H. von Lohneysen, T.Pietrus, G.Portisch, H.G.Schlager, A.Schroder, M.Sieck, T.Trappmann, Phys.Rev.Lett., **72**, 3262(1994).

2. B.Bogenberger, H. von Lohneysen, Phys.Rev.Lett., **74**, 1016(1995).

3. A.Rosch, A.Schroder, O.Stockert, H. von Lohneysen, Phys.Rev.Lett., **79**, 159(1997).

4. A.Schroder, G.Aeppli, E.Bucher, R.Ramazashili, P.Coleman, Phys.Rev.Lett., **80**, 5629(1998).

5. O.Stockert, H. von Lohneysen, A.Rosch, N.Pyka, M.Loewenhaupt, Phys.Rev.Lett., **80**, 5627(1998).

6. H. von Lohneysen, C.Pfleiderer, T.Pietrus, O.Stockert, B.Will, Phys.Rev.B, **63**, 134411(2001).

7. O.Stockert, F. Huster, A.Neubert, C.Pfleiderer, T.Pietrus, B.Will, H. von Lohneysen, Physica B, **312-313**, 458(2002).

8. G.S.Stewart, Z.Fisk, M.S.Wire, Phys. Rev. B, **30**, 482 (1984).

9. K.Andres, J.E.Graebner, H.R.Ott, Phys. Rev. Lett., **35,** 1779 (1975)

10. H. von Lohneysen, A.Neubert, T.Pietrus, A.Schroder, O.Stockert, U. Tutsch, M.Loewenhaupt, A.Rosch, P.Wolfle, Eur. Phys. J. B, **5**, 447(1998).

11. P.Coleman, C.Pepin, Q.Si, R.Ramazashvili, J.Phys.: Cond. Mat., **13**, R723(2001).

12. T.Penney, J.Stankiewicz, S. von Molnar, Z.Fisk, J.L.Smith, H.R.Ott, J. Magn.Magn.Mat., **54-57**, 370(1986).

13. T.Penney, F.M.Milliken, S. von Molnar, F.Holtzberg, Z.Fisk, Phys. Rev. B, **34**, 5959(1986).

14. Y.Onuki, Y.Shimizu, M.Nishihara, Y.Machii, T.Komatsubara, J. Phys. Soc. Jpn., **54**, 1964(1985).

15. Y.Onuki, T.Yamazaki, T.Omi, I.Ukon, A.Kobori, T.Komatsubara, J. Phys. Soc. Jpn., **58**, 2126(1989).

16. T.Namiki, H.Sato, J.Urakawa, H.Sugawara, Y.Aoki, R.Settai, Y.Onuki, Physica B, **281-282**, 359(2000).

17. H.Bartolf, C.Pfleiderer, O.Stockert, M.Vojta, H. von Lohneysen, Physica B, **359-360**, 86(2005).

18. N.E.Sluchanko, A.V.Bogach, V.V.Glushkov, S.V.Demishev, M.I.Ignatov, N.A.Samarin, G.S.Burkhanov, O.D.Chistyakov, JETP, **98**, 793(2004).

19. N.E.Sluchanko, A.V.Bogach, G.S.Burkhanov, O.D.Chistyakov, V.V.Glushkov, S.V.Demishev, N.A.Samarin, D.N.Sluchanko, Physica B, **359-361,** 308(2005).

20. Y.Onuki, Y.Furukawa, T.Komatsubara, J. Phys. Soc. Jpn., **53**, 2197(1984).

21. J.Rossat-Mignod, L.P.Regnault, J.L.Jacoud, C.Vettier, P.Lejay, J.Flouquet, E.Walker, D.Jaccard, A.Amato, J.Magn.Magn.Mat., **76-77**, 376(1988).

22. H. von Lohneysen, M.Sieck, O.Stockert, M.Waffenschmidt, Physica B, **223-224**, 471(1996).

23. N.E.Sluchanko, V.V.Glushkov, S.V.Demishev, N.A.Samarin, G.S.Burkhanov, O.D.Chistyakov, D.N.Sluchanko, cond-mat/0505386.

24. P.Coleman, P.W.Anderson, T.V.Ramakrishnan, Phys. Rev. Lett., **55** (1985) 414.

25. A.Fert, P.M.Levy, Phys. Rev. B, **36,** 1907 (1987).

26. E.A.Goremychkin, R.Osborn, Phys. Rev. B, **47**, 14580 (1993).

27. V.V.Glushkov, I.B.Voskoboinikov, S.V.Demishev, I.V.Krivitskii, A.Menovsky, V.V.Moshchalkov, N.A.Samarin, N.E.Sluchanko, JETP, **99**, 394 (2004).

28. N.E.Sluchanko, V.V.Glushkov, B.P.Gorshunov, S.V.Demishev, M.V.Kondrin, A.A.Pronin, A.A.Volkov, A.K.Savchenko, G.Gruner, Y.Bruynseraede, V.V.Moshchalkov, S.Kunii, Phys.Rev.B, **61**, 9906 (2000).

**Figure Captions.**

**Fig. 1.** Temperature dependencies of resistivity $\rho(T)$ $CeCu_{6-x}Au_x$ (x=0, 0.1, 0.2) for various values of magnetic field (H=0 kOe, 70 kOe).

**Fig. 2.** Angular dependencies of the Hall resistance $\rho_H(\varphi, T_0)$ of $CeCu_6$ (a) and $CeCu_{5.8}Au_{0.2}$ (b) in magnetic field $H_0 \approx 3.7$ kOe at different temperatures.

**Fig. 3.** Angular dependencies of the Hall resistance of $CeCu_{5.9}Au_{0.1}$ at various temperatures in magnetic field $H=3.7$ kOe and separation of contributions to (1) (see text); $\rho_H^{exp}$ are experimental data and fit by (1), $\rho_{H1}$ is the main component in the Hall signal and $\rho_{H2}$ is the second harmonic contibution.

**Fig. 4.** Temperature dependence of the Hall coefficient of $CeCu_{6-x}Au_x$ (x=0, 0.1, 0.2) in logarithmic coordinates. Inset shows the scheme of crystal field splitting of cerium $^2F_{5/2}$-state.

**Fig. 5.** Temperature dependecies of the second harmonic term $R_{H2}(T)$ in the Hall effect of $CeCu_{6-x}Au_x$. Inset shows a reciprocal plot $R_{H2} = f(1/T)$.

**Fig. 6.** Temperature dependence of the effective reduced concentration of carriers per cerium site $1/(n_{4f}eR_H)$ of $CeAl_3$ and $CeCu_6$ in reciprocal logarithmic coordinates in the intervals I, II (see text); $n_{4f}(CeAl_3)=6.9 \cdot 10^{21} cm^{-3}$, $n_{4f}(CeCu_6)=9.5 \cdot 10^{21} cm^{-3}$.

**Fig. 7.** Temperature dependencies of the Hall mobility $\mu_H(T) = R_H(T)/\rho(T)$ in $CeAl_3$ and $CeCu_{6-x}Au_x$ (x=0, 0.1, 0.2). The linear fits correspond to the power-law behavior of $\mu_H(T) \sim T^{-\alpha}$. Inset shows the temperature dependencies of the phase shift between harmonics in $CeCu_6$ and $CeCu_{5.9}Au_{0.1}$ (see text).

**Table 1.** The lattices constants in $CeCu_{6-x}Au_x$.

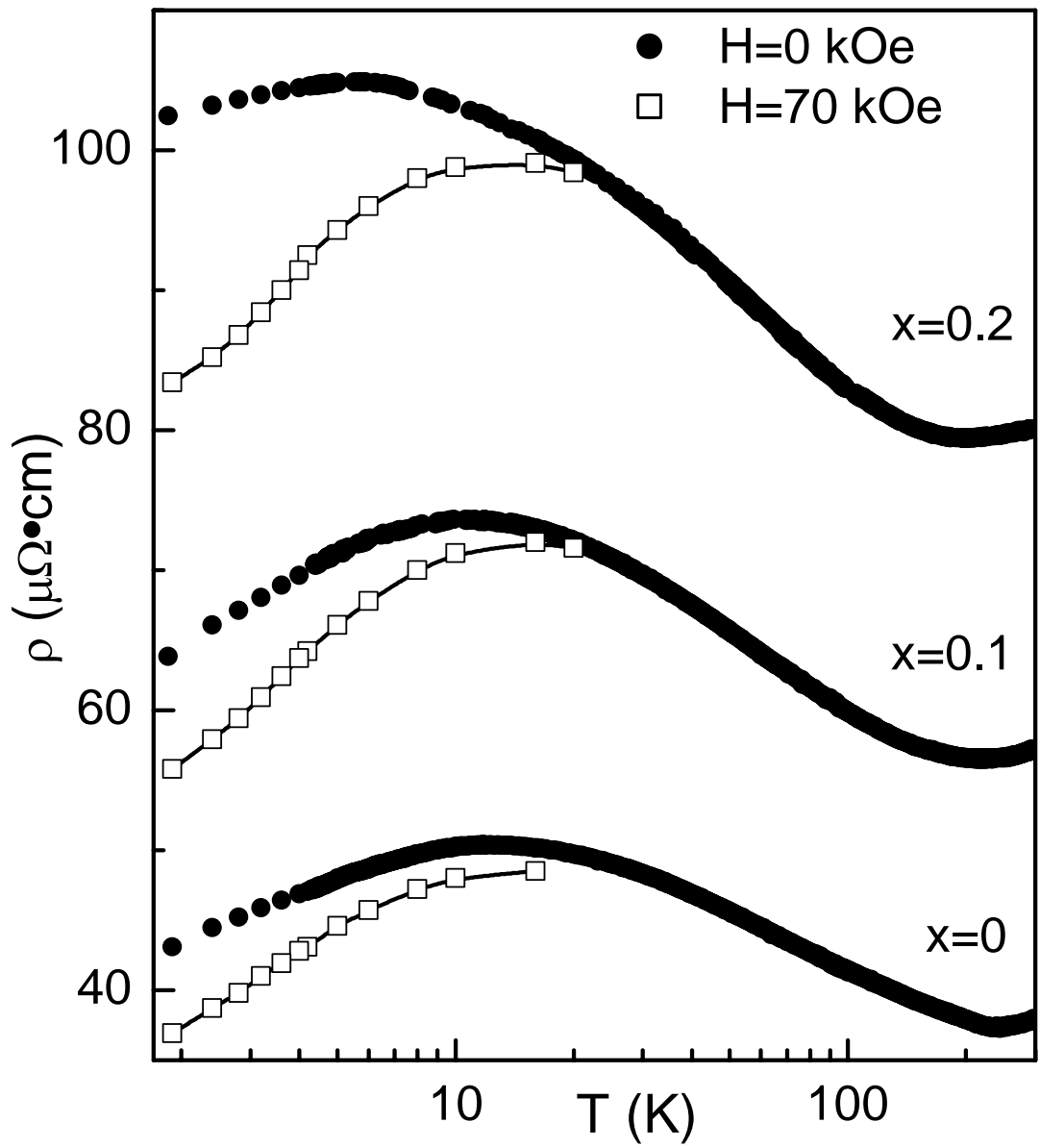

Fig.1

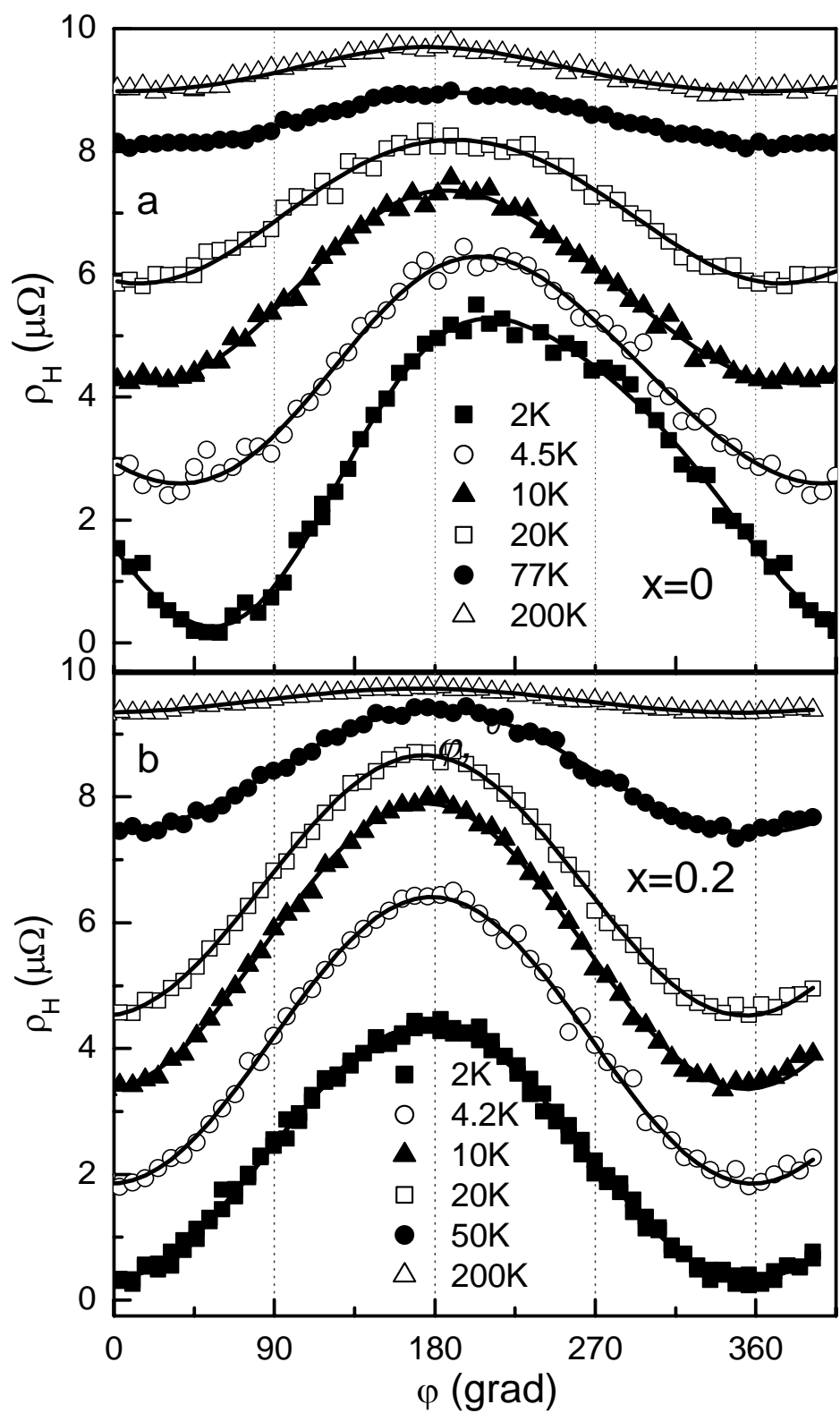

Fig.2

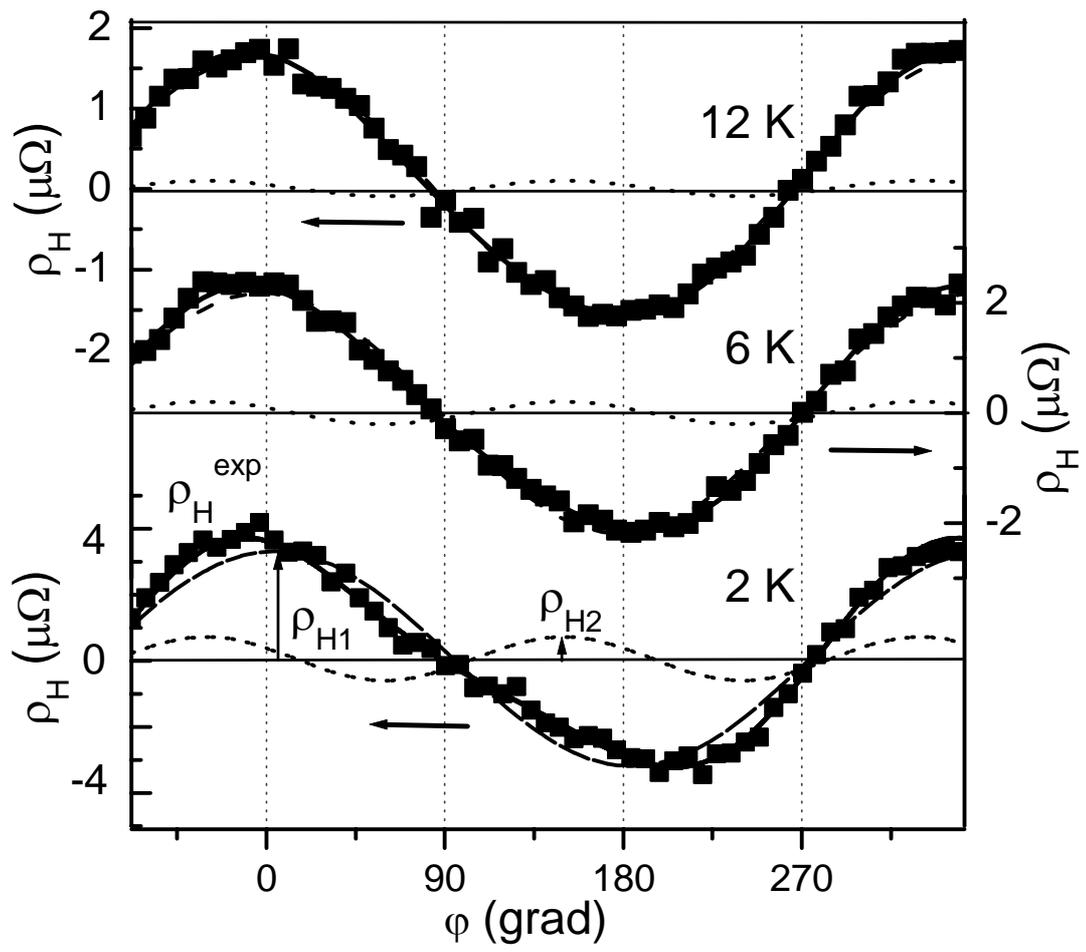

Fig.3

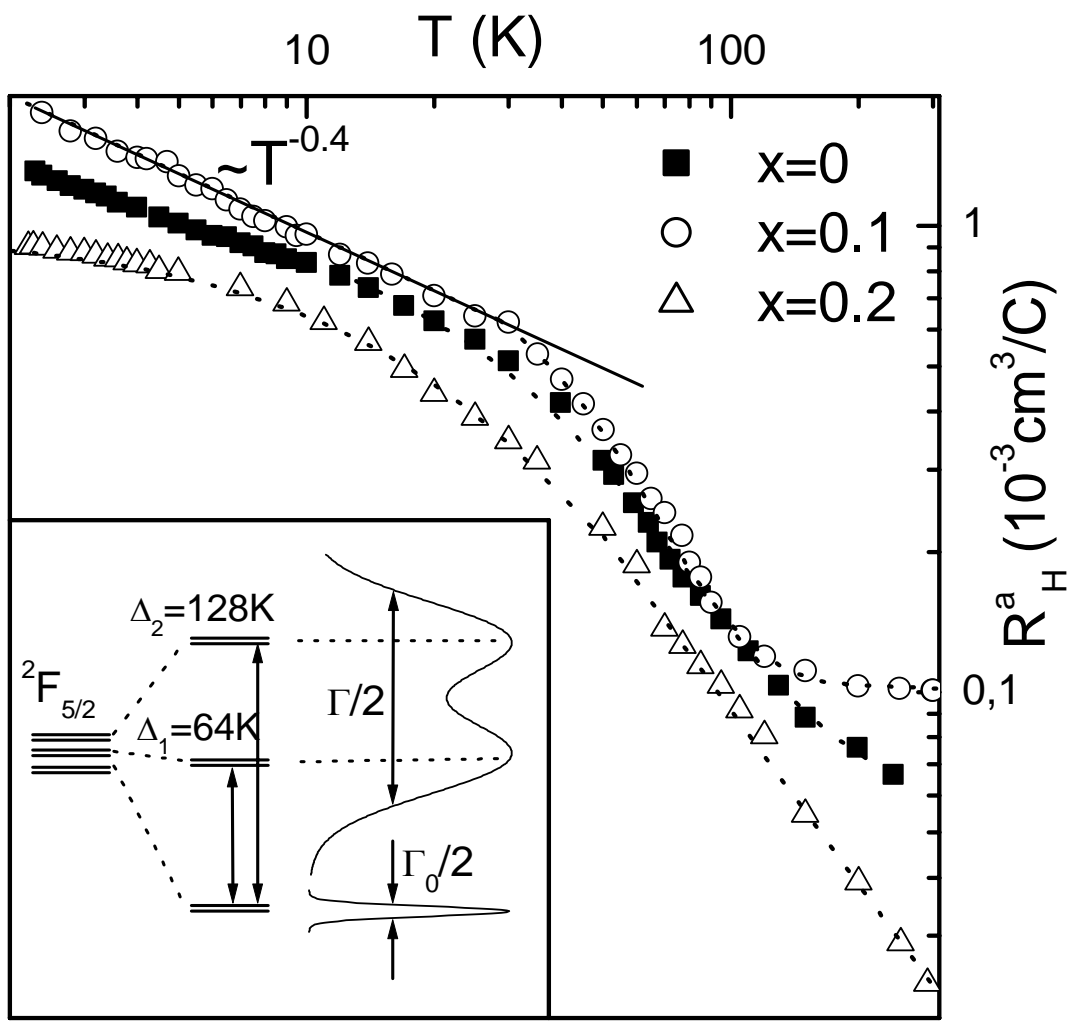

Fig.4

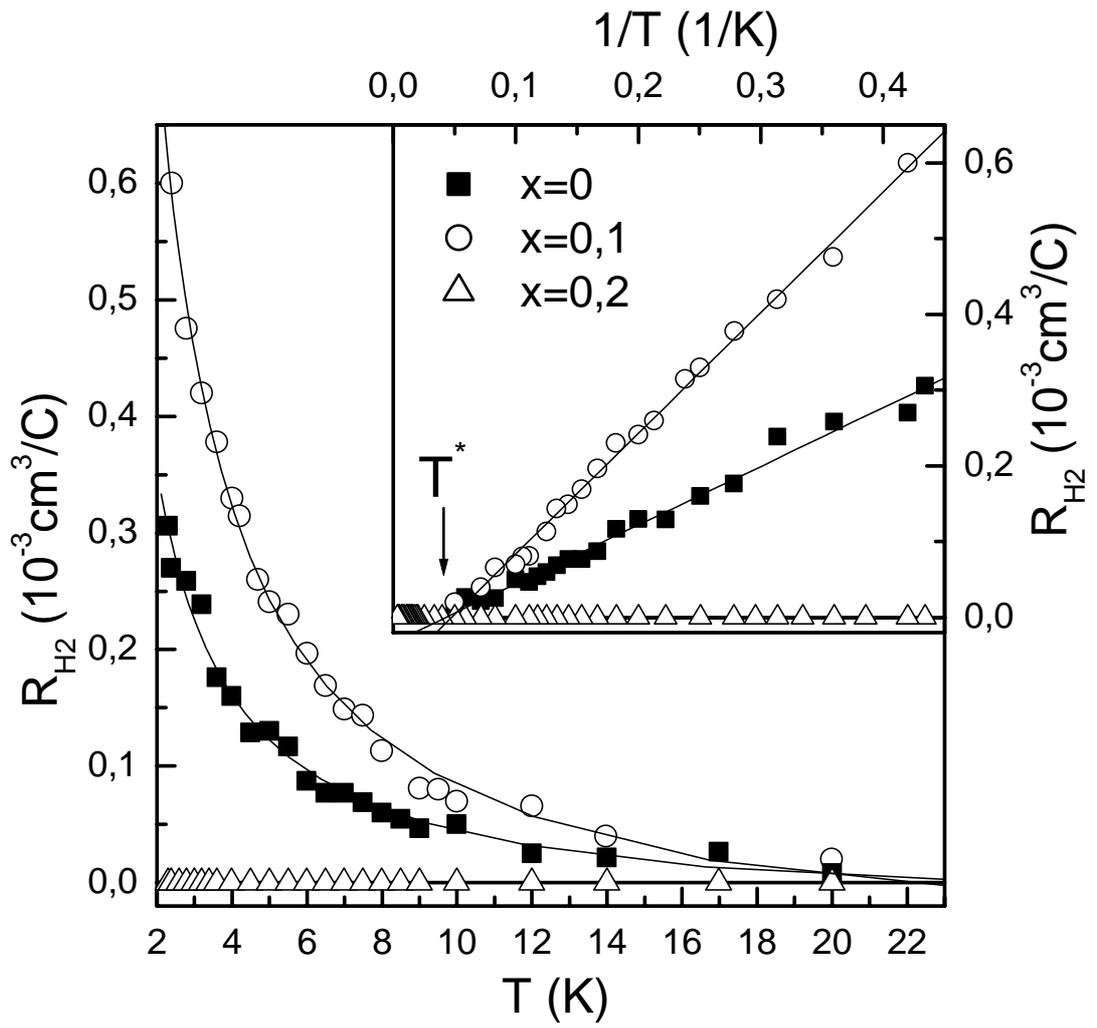

Fig.5

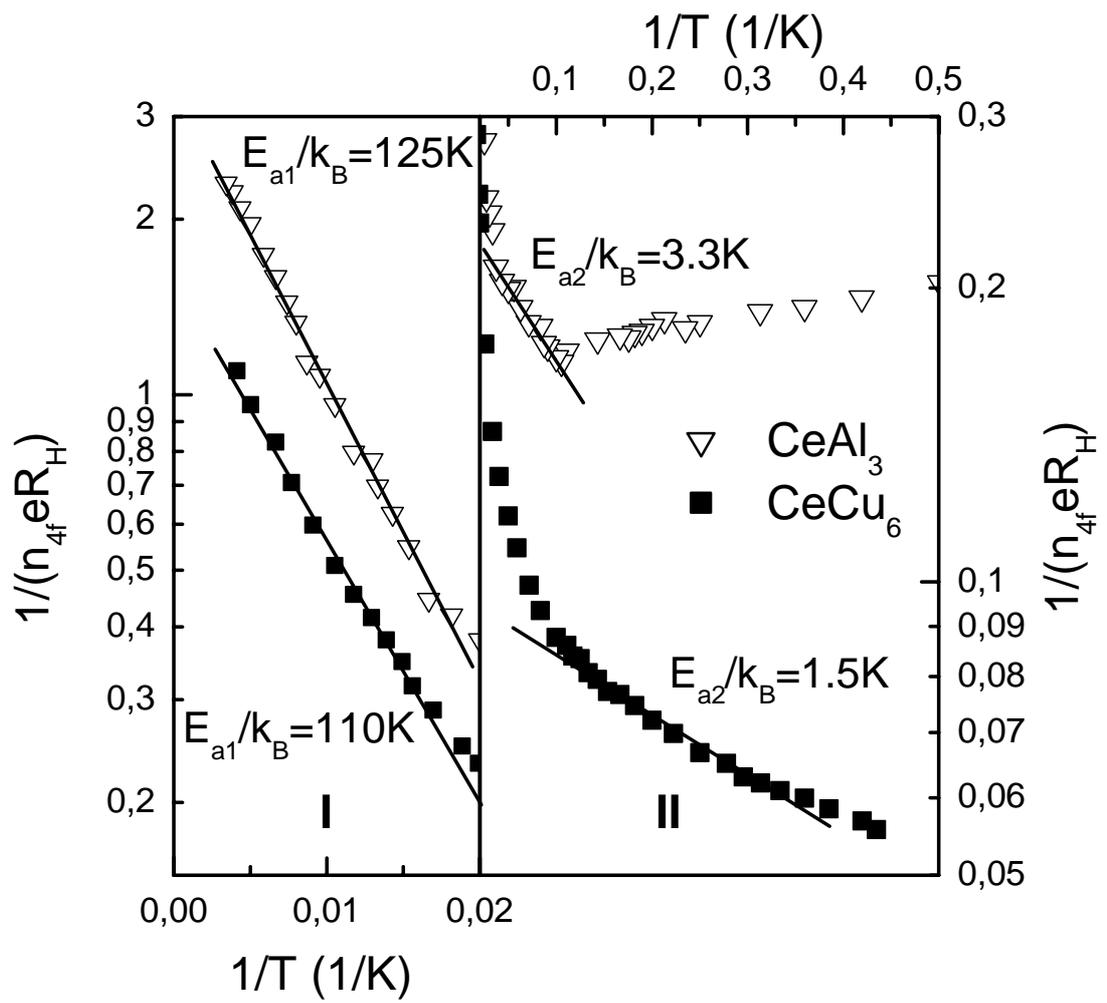

Fig.6

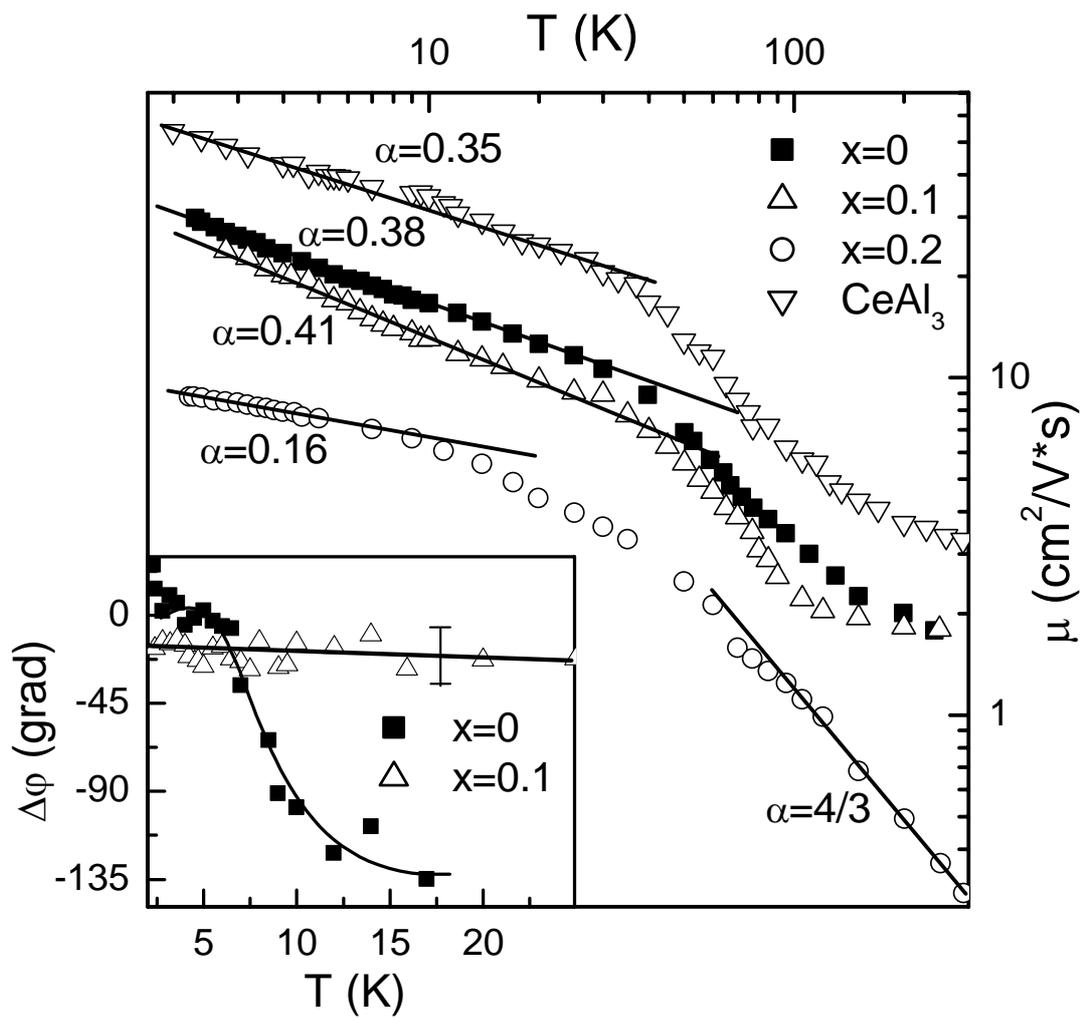

Fig.7

**Table 1.**

| x | a, Å | b, Å | c, Å | $v_l$, Å / $\beta$, grad |
|---|---|---|---|---|
| 0 (orthorh.) | 8.108(3) | 5.103(5) | 10.163(5) | 420.51(50) |
| 0.1 (orthorh.) | 8.109(3) | 5.102(2) | 10.162(2) | 420.41(40) |
| 0.2 (orthorh.) | 8.115(2) | 5.092(2) | 10.177(2) | 420.47(39) |
| 0 (orthorh.) [20] | 8.109 | 5.105 | 10.159 | 420.55 |
| 0 (monocl.) [20] | 5.080 | 10.121 | 8.067 | $\beta$=91.36° |